\documentclass[%
 reprint,
 amsmath,amssymb,
 aps,
prl,
]{revtex4-2}

\usepackage{graphicx}
\usepackage{dcolumn}
\usepackage{bm}
\begin{document}

\preprint{APS/123-QED}

\title{All-Dielectric Metasurface with a Two-Dimensional Locally Flat Photonic Band}

\author{Minho Choi}
\thanks{These authors contributed equally to this work: Minho Choi, Christopher Munley and Virat Tara.}
 \affiliation{
 Department of Electrical and Computer Engineering, University of Washington, Seattle, WA 98195, USA
}

\author{Christopher Munley}
\thanks{These authors contributed equally to this work: Minho Choi, Christopher Munley and Virat Tara.}
 \affiliation{
 Department of Physics, University of Washington, Seattle, WA 98195, USA
} \affiliation{
 Laboratory for Physical Sciences, College Park, MD 20740, USA
}

\author{Virat Tara}
\thanks{These authors contributed equally to this work: Minho Choi, Christopher Munley and Virat Tara.}
 \affiliation{
 Department of Electrical and Computer Engineering, University of Washington, Seattle, WA 98195, USA
}

\author{Arnab Manna}
 \affiliation{
 Department of Physics, University of Washington, Seattle, WA 98195, USA
}

\author{Johannes Fr\"och}
 \affiliation{
 Department of Electrical and Computer Engineering, University of Washington, Seattle, WA 98195, USA
}

\author{Arthur Barnard}
 \affiliation{
 Department of Physics, University of Washington, Seattle, WA 98195, USA
}

\author{Arka Majumdar}
 \email{arka@uw.edu}
 \affiliation{
 Department of Electrical and Computer Engineering, University of Washington, Seattle, WA 98195, USA
}
\affiliation{
 Department of Physics, University of Washington, Seattle, WA 98195, USA
}

\date{\today}

\begin{abstract}
Photonic flatbands offer promising light-matter interaction due to their unique slow-light nature. In recent years, flatbands have also attracted significant interest in optical engineering because of their angle-insensitive resonant characteristics. However, to date, no studies have reported the dispersionless behavior of flatbands under arbitrary two-dimensional incident angles and polarizations. Here, we present a two-dimensional photonic flatband created using a silicon metasurface with a Lieb lattice-inspired structure which demonstrates a locally flat photonic band for both transverse electric (TE) and transverse magnetic (TM) polarized light. Employing Fourier imaging, we analyze the energy-momentum dispersion of the flatband metasurface under arbitrary two-dimensional incident angles, demonstrating dispersionless flatbands with a change in resonance within $\pm2$ nm up to $\pm24^\circ$ or $\pm5^\circ$, depending on the polarization state and incident angle. This geometry can be adapted for various applications in local field enhancement, enhanced photodetection, and augmented reality displays.

\end{abstract}
\maketitle
The propagation of light without dispersion in photonics has historically been studied under the topic of \textit{slow} light, mostly in the context of photonic crystals \cite{baba2008slow} and coupled waveguide arrays \cite{mukherjee2015observation,vicencio2015observation}. The dispersionless photonic band structure, also referred to as a flatband, has zero group velocity, resulting in highly localized modes \cite{caceres2019perfect}. Electric flatbands have generated significant interest due to their potential to demonstrate exotic phenomena such as quantum scarring \cite{kuno2020flat} and Aharonov-Bohm caging \cite{vidal2001disorder,vidal1998aharonov,li2022aharonov}. On the other hand, photonic bands which are only locally flat can increase the photonic density of state and light-matter interaction \cite{hau1999light} potentially leading to strong optical nonlinearity \cite{krauss2008we,hail2024third,jiang2025flatband}. With the advent of sub-wavelength patterning of dielectric thin films, photonic flatbands have been demonstrated in various structures including twisted bilayer moir\'e superlattices \cite{tang2021modeling,raun2023gan,jing2025observation}, guided mode resonators \cite{nguyen2018symmetry,munley2023visible}, and two-dimensional coupled photonic crystal cavity arrays \cite{altug2005experimental,yang2017topological}.

In recent years, photonic flatbands have showcased significant potential in various engineering applications, including photodetectors \cite{choi2024nonlocal}, lasers \cite{raun2023gan,eyvazi2025flat,cui2025ultracompact}, and free-electron radiation sources \cite{yang2023photonic}, owing to their angle-independent optical responses. However, in most photonic structures, flatbands are confined to one spatial dimension and sensitive to one specific linear polarization. Notably, recent advancements have extended the applicability of flatbands to circularly polarized light \cite{choi2025observation}. Despite this progress, the use of photonic flatbands in free space imaging applications has been limited by a lack of demonstrated two-dimensional, polarization-independent, flatband behavior \cite{tang2021low}.

Generally, light scattering by spherical particles is independent of the incident light’s direction or state of polarization due to their inherent symmetry \cite{sinclair1947light}. Similarly, plasmonic cavities exhibit comparable insensitivity to the angle of incidence. Highly localized surface plasmonic modes, along with their short-range interactions within individual structures, reduce dependence on structural configurations \cite{fan2014light}. Additionally, the broadband response of plasmonic modes typically shows low sensitivity to the angle of incident light \cite{ye2015angle,mao2016angle,wells2021angle}. However, plasmonic structures face limitations as practical optical components due to inevitable metallic absorption $-$ resulting in quality factors constrained to below 100 $-$ and their relatively broad spectral responses. In contrast, dielectric photonic structures, such as nonlocal metasurfaces, can achieve narrow bandwidths with minimal optical loss \cite{overvig2022diffractive,huang2023ultrahigh}. Nonetheless, the long-range interactions characteristic of dielectric metasurfaces typically render their guided modes highly sensitive to the direction and polarization of incident light. Consequently, achieving photonic flatbands across all polarizations and angles of incidence remains a significant challenge for nonlocal dielectric metasurfaces.

In this paper, we employ a variant of Lieb lattice, sometimes referred to as a decorated square lattice (a square lattice with missing site), to create dielectric metasurfaces exhibiting two-dimensional photonic flatbands. The Lieb lattice based photonic flatband is first realized by evanescently coupled waveguides using laser writing techniques \cite{mukherjee2015observation,vicencio2015observation}. And it garnered attention due to its remarkable robustness to perturbations \cite{molina2015flat} and as an unconventional simulation platform according to its dispersionless and extremely localized states. Distortion-free image transmission has been reported employing its angle-insensitive response \cite{xia2016demonstration}. Exciton-polariton condensations have been observed in (Al,Ga)As systems with Lieb lattice configurations \cite{klembt2017polariton,whittaker2018exciton,harder2020exciton}. However, comprehensive studies examining two-dimensional flatbands in relation to various incident light’s polarization angles remain absent. In fact, a straightforward implementation of Lieb lattice in metasurface (unit cell containing one missing hole in a 2x2 array of holes) does not provide a flatband (See Supplementary Figure 1). As such, to the best of our knowledge, two-dimensional photonic flatbands that are locally flat across all polarizations and angles of incidence have yet to be demonstrated.

We design and fabricate a silicon metasurface based on a Lieb lattice-inspired photonic crystal with broken vertical symmetry, demonstrating a locally flat photonic band with a moderately high quality factor ($\sim$300). We choose silicon for its high refractive index, which enables guided modes to exist within a thin metasurface membrane while minimizing the emergence of unwanted higher-order modes. We emphasize that while high index of the material is a necessary condition to observe the flatband, it is not sufficient on it own to create a flatband resonator. We also note that our structure is not extremely sensitive to index variation, and as other materials with high index, such as Gallium Phosphide will also allow us to achieve flatband \cite{munley2023visible}. To analyze the dispersion of the photonic band structure, we employ energy-momentum spectroscopy across various angular orientations and polarizations.

We start our design of a flatband from a Lieb lattice of circular holes, which did not produce the desired flatband. By leveraging the symmetries of the lattice and precisely fine-tuning the unit cell elements, the photonic bands are engineered to remain flat for all polarizations and directions within the Brillouin zone near the $\Gamma$-point, up to a numerical aperture of 0.1 (equivalent to $\pm$5$^o$ of incident angles). We found the eccentrities of the holes to be an important parameter to ensure polarization-independent two-dimensional flatband. Rigorous coupled wave analysis (RCWA) shows excellent agreement with experimentally measured energy-momentum dispersion of the flatband.

\begin{figure}[b]
\includegraphics{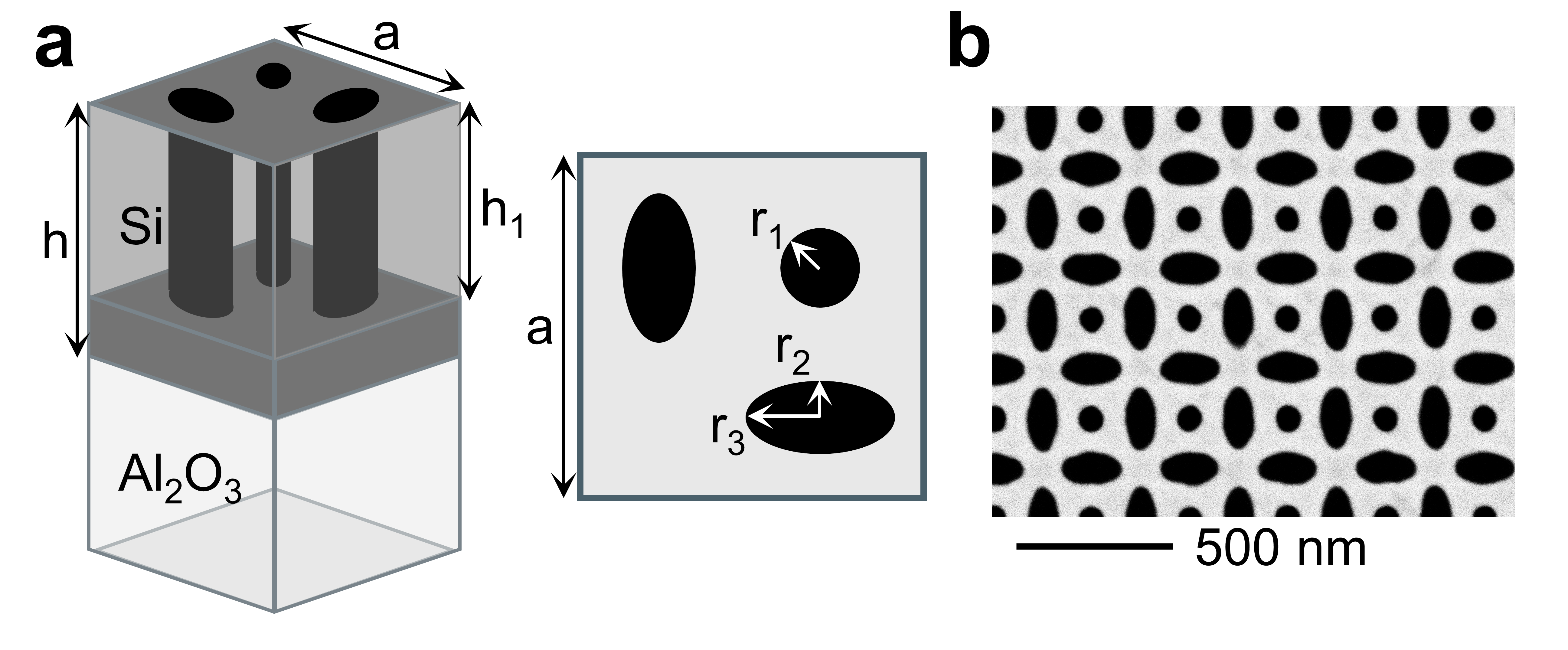}
\caption{\label{fig:Figure1}Flatband metasurface geometry. (a) (left) Crystalline silicon of thickness, h = 230 nm, on a sapphire substrate highlighting the partial etch of Mie voids into the crystalline silicon layer down to a depth h\textsubscript{1} = 223 nm. (right) Diagram of the designed metasurface Lieb lattice unit cell of period, a = 309 nm, showing both the major and minor sublattices separated by a spacing, a/2. The minor sublattice is composed of circular voids with radius, r\textsubscript{1} = 42.5 nm, and the major sublattice is composed of elliptical voids with semi-minor axis, r\textsubscript{2} = 42.5 nm and semi-major axis, r\textsubscript{3} = 66.5 nm. (b) Scanning electron micrograph of the fabricated crystalline silicon metasurface.}
\end{figure}

The simulation and design of the metasurface are conducted by the Lumerical RCWA, which functions similarly to the S4 solver, enabling the analysis of photonic band structures \cite{liu2012s4}. Figure \ref{fig:Figure1}a shows a unit cell of the metasurface. By tuning the partial etching parameter, h\textsubscript{1}, in a vertical-symmetry-broken Lieb lattice photonic crystal, a photonic band that is flat for one polarization and one crystal axis can be obtained \cite{nguyen2018symmetry}. Once the band becomes flat for a given polarization and principal axis in the Brillouin zone, this behavior can be extended to other polarizations through modification of the lateral parameters while preserving the initially flatband behavior. The fourfold symmetry of the Lieb lattice naturally extends the engineered flatband to both principal axes of the lattice, but by disrupting the rotational symmetry of the major sublattice elements and reshaping them into ellipses (Fig. \ref{fig:Figure1}a), the band curvature can be selectively tuned to a flatband for both S- and P-polarizations. This is due to the fact that when the S-polarized and P-polarized bands are degenerate at the $\Gamma$-point, each polarization is seen in simulation to correspond to a field primarily confined to one of the major sublattice elements of the unit cell (Supplementary Figure 1). Thus, a change in the eccentricity of the major sublattice elements can create a differential shift in energy between S-polarized and P-polarized bands at various incident angles and can be used to effect a near flat dispersion for both bands. Further increases in the eccentricity of the major sublattice elements ultimately cause the band curvature to deviate from a flat profile (Supplementary Figure 2). With engineered flatband dispersion for a given span along the principal axes of the lattice, and for both S-polarized and P-polarized incident light, simulation of the Lieb lattice metasurface shows flatband dispersion along various azimuthal axes including along the $\Gamma$-M axis.

With the flatband identified and fine-tuned to span all axes and polarizations, the focus shifts to accurately fabricating the design. The metasurface is constructed using a 230-nm-thick crystalline silicon film on a sapphire (Al\textsubscript{2}O\textsubscript{3}) substrate. Electron beam lithography, followed by reactive ion etching with combined fluorine gases (SF\textsubscript{6}/C\textsubscript{4}F\textsubscript{8}), facilitates the creation of the all-dielectric metasurface as shown in Fig. \ref{fig:Figure1}b. The partial etching depth is controlled through the etching time in the absence of an etch-stop layer. Fabrication imperfections, such as inaccurate partial etch depth, etched sidewall roughness, and excessive pattern enlargement due to overexposure or overetching, lead to unexpected variations in the simulated parameters. To address these challenges, the dependence of band bending on the partial etch depth, h\textsubscript{1}, is systematically analyzed. Despite the calibration provided by test etching, errors in partial etch depth exert a considerable influence on band curvature. Consequently, separate lateral parameters ($a$ and $a/r_i$) are swept across an array of devices to ensure that at least one device successfully achieves the desired flatband.

\begin{figure*}[htbp]
\includegraphics{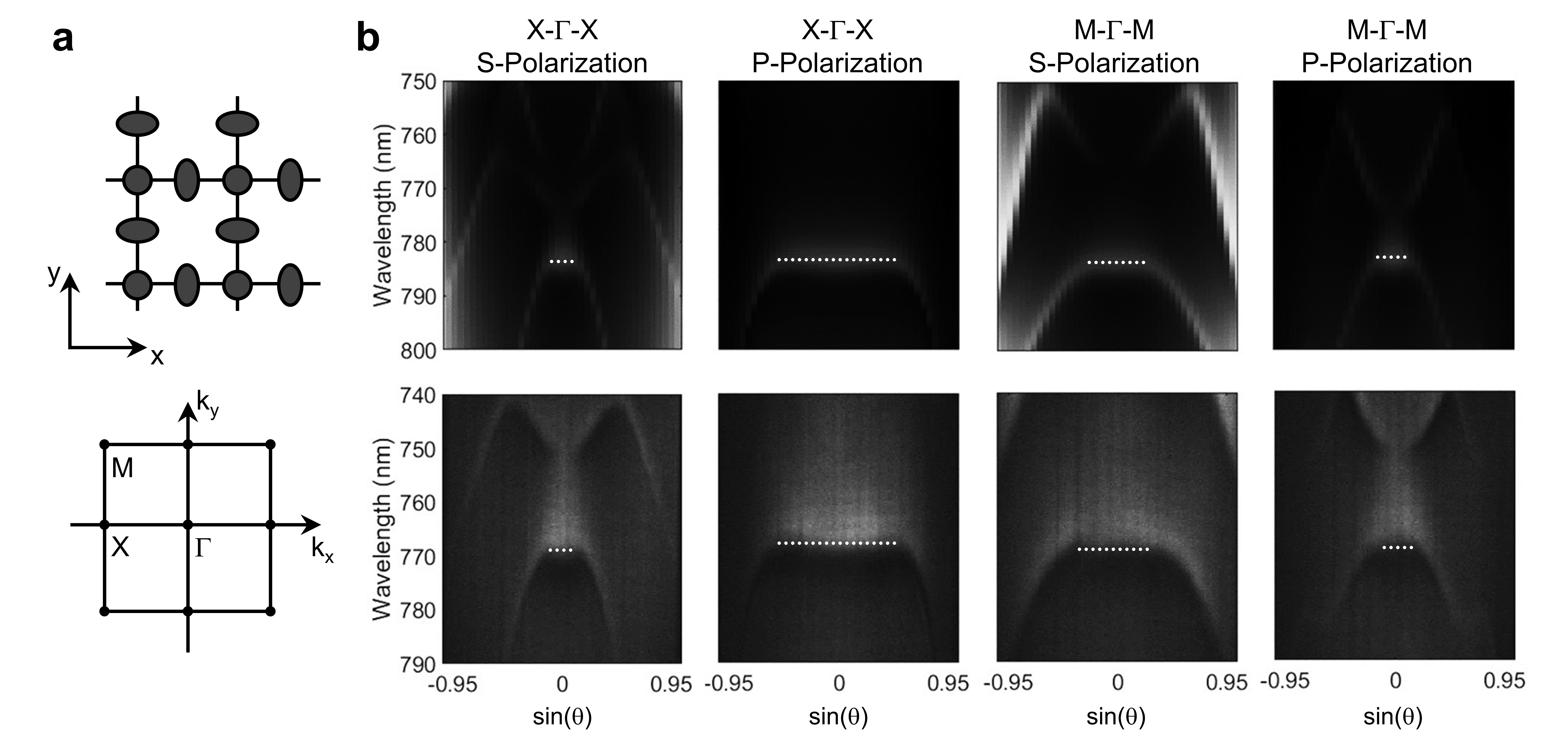}
\caption{\label{fig:Figure2} (a) Real space and first Brillioun zone representations of the Lieb lattice. (b) Comparison of simulated band structure (top row) with experimentally measured band structure (bottom row) for the Lieb lattice metasurface. Band structures up to the numerical aperture of 0.95 for the objective are shown along the X and M axes in the Brillouin zone for both S- and P-polarizations. White dashed lines indicate flatbands at each polarizations and oblique incident directions. Note a wavelength offset of 10nm between the simulation and experiment, which is most likely due to fabrication imperfection. A change in resonance wavelength within $\sim$ $\pm$ 2 nm was considered to be within the flatband extent.}
\end{figure*}
 
We measure the energy-momentum dispersion of the two-dimensional flatband using Fourier imaging. A minimal entrance slit width is employed to filter along the selected axis of the Brillouin zone. This combination of imaging and filtering, together with the diffractive spread of the spectrometer, projects the band structure onto the spectrometer camera, capturing the reflection of the metasurface as a function of wavelength and angle of incidence. To enable direct comparison, we generate RCWA simulations of the reflection coefficient using the same measurement parameters. By physically rotating the metasurface around the optical axis of the objective, the back focal plane image rotates relative to the entrance slit of the spectrometer, facilitating measurements along different principal axes of the Brillouin zone. Inserting a polarizer directly before the spectrometer allows for the measurement of different polarizations while minimizing polarization crosstalk introduced by preceding optics. Illumination is achieved by imaging the fiber-coupled output of a stabilized tungsten-halogen source through a pinhole, effectively conditioning the illumination field for the energy-momentum spectroscopy. See the schematic for an experimental setup in an Supplementary Figure 3.

As shown in Figure \ref{fig:Figure2}, the measured energy-momentum dispersion of the flatband aligns closely with the simulated dispersion. However, a wavelength shift of approximately 10 nm is observed in the resonance of the photonic bands between the simulations and measurements. Structural analysis reveals a partial etch depth of 223 nm, which exceeds the expected depth of 209 nm by 14 nm. Simulations investigating this partial etch error suggest that such over-etching would typically drive the photonic band toward a parabolic dispersion (Supplementary Figure 4). However, the measured flatband dispersion indicates that compensatory adjustments in lateral parameters most likely counteract this effect, maintaining the flatband condition.

\begin{figure*}[htbp]
\includegraphics{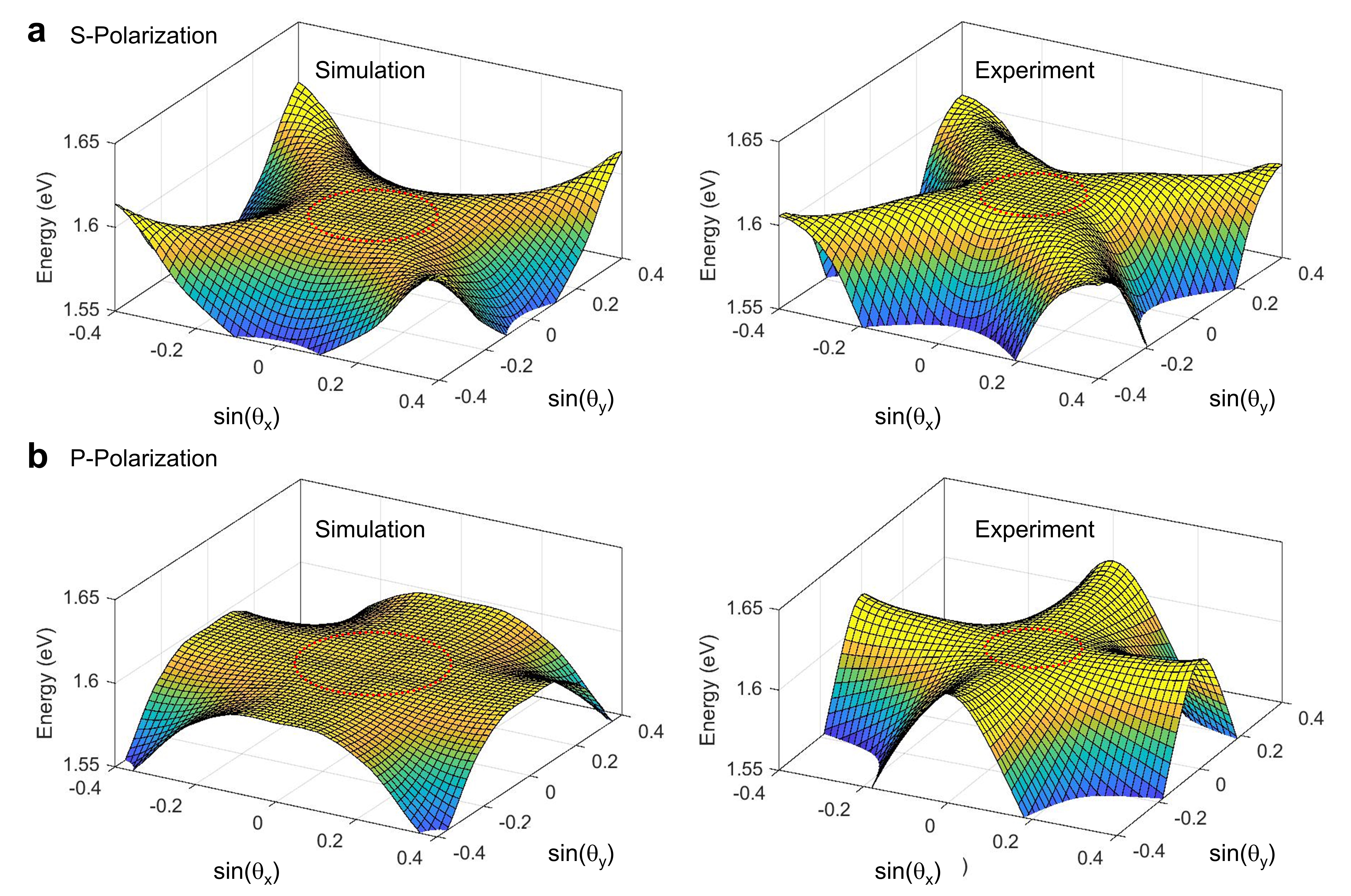}
\caption{\label{fig:Figure3}Extracted bands for simulated and measured S-polarized light as well as for simulated and measured P-polarized light (a and b, respectively). Band surfaces are extracted by fitting Fano resonances to spectra taken at 15$^\circ$ increments in azimuthal and applying a smoothing filter to interpolate a continuous band surface. Results are shown in the vicinity of the $\Gamma$-point up to a numerical aperture of 0.4. While the S-polarized band shows remarkable agreement between simulation and experiment (a), the P-polarized band is more clearly flat in simulation than in experiment (b). Red dashed lines indicate flatband regions at each polarizations. }
\end{figure*}

In addition, we measure the energy-momentum dispersion of the flatband in arbitrary directions of incident angle by rotating the sample in 15$^\circ$ increments. With the slit on the spectrometer fixed, we can selectively collect specific one-dimensional energy-momentum dispersion data. A smoothing filter in MATLAB software is applied to interpolate a continuous band surface. Simultaneously, we characterize the polarization dependence of the band structure for both S- and P-polarization directions. Figure \ref{fig:Figure3} presents the simulated and measured energy-momentum dispersion of the all-dielectric metasurface in the vicinity of the $\Gamma$-point, extending up to a numerical aperture of 0.4 (within an angle of incidence of $~24^o$). The experimentally observed two-dimensional band structure aligns well with the simulation, though a $\sim$10 nm shift in the resonant wavelength is noted. The symmetries and structure of the bands are consistent; however, the experimentally measured dispersion for P-polarized light, while maintaining a flat portion near the $\Gamma$-point, exhibits less flat dispersion farther along the $\Gamma$-X axis.

\begin{figure}[htbp]
\includegraphics[width=\columnwidth]{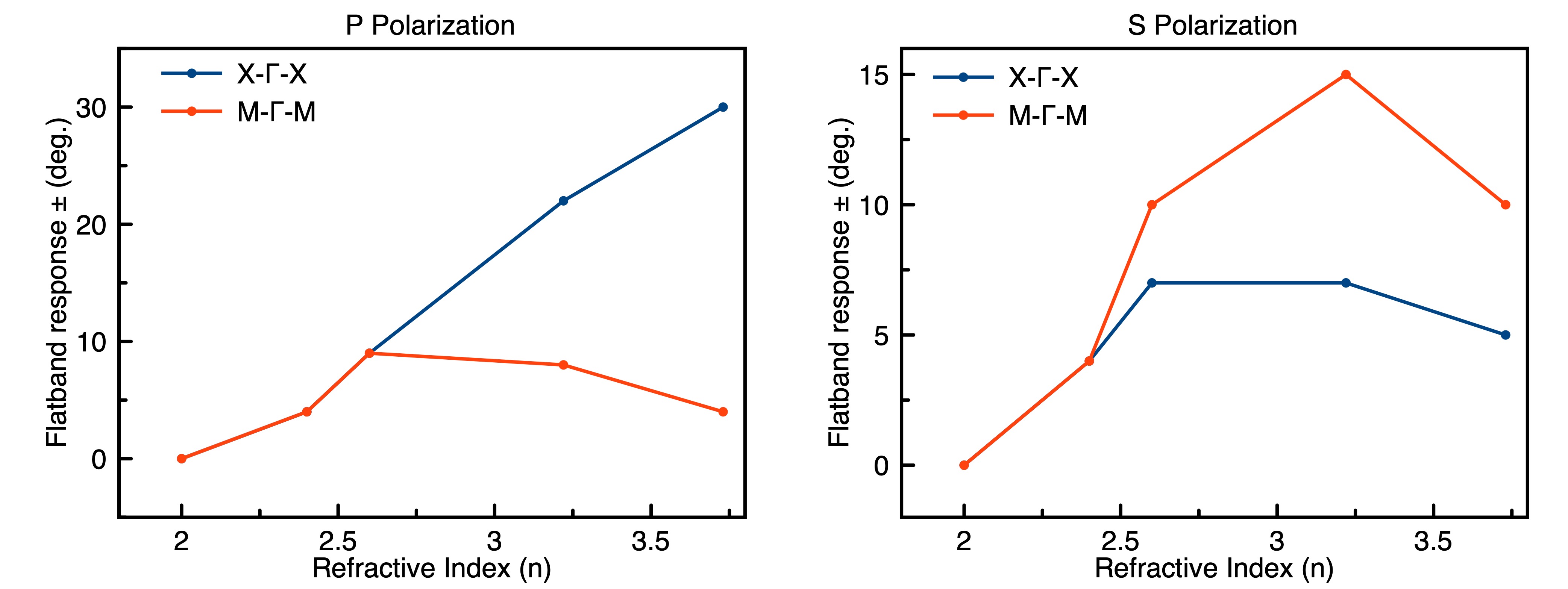}
\caption{\label{fig:Figure4}Flatband response versus refractive index of the metasurface for (a) P- and (b) S- polarizations. A change in resonance wavelength within $\sim$ $\pm$ 2 nm was considered to be within the flatband extent.}
\end{figure}

Next, we also explore the effect of the refractive index (n) on the response of our 2D flatband metasurface. We study the change in response of the metasurface if we use materials such as Gallium Phosphide (GaP) $n=3.22$, Silicon Carbide (SiC) $n=2.6$, Titanium Dioxide (TiO\textsubscript{2}) $n=2.4$, Silicon Nitride $n=2.0$ (Si\textsubscript{3}N\textsubscript{4}) instead of Si. From simulations we note that a decrease in the refractive index adversely affects the range of the flatband, as can be seen more evidently for p-polarization along X-$\Gamma$-X as seen in Figure \ref{fig:Figure4}. No flatband was observed for $n=2.0$. To keep the resonance wavelength roughly at ~800 nm wavelength, same as our original design, we scale all the geometery parameters of the metasurface shown in Figure \ref{fig:Figure1}  by a fraction of $n_{si}/n$ where $n_{si}$ was set as $3.73$. The unetched thickness for Silicon metasurface was kept as 209 nm and Sapphire with $n=1.76$ was used as the substrate material.  To calculate the extent of the flatband, a change in resonance wavelength within $\sim$ $\pm$2 nm was considered to be within the flatband extent.

During the past decade, significant advances have been made in various meta-optics, demonstrating remarkable performance as filters and free-space optical cavities \cite{tang2024metasurface}. However, in most cases, meta-optics are extremely prone to misalignment, and their functionality (as wavelength/ polarization filters or as resonators) are highly sensitive to the angles or wavefronts of incident light. For instance, the cutoff wavelength of a meta-optic filter can vary depending on the incident angle of the light. This sensitivity, known as an off-axis aberration, often forces a compromise between system performance and robustness \cite{aieta2013aberrations}.

In this work, we address this seemingly inevitable trade-off by implementing a two-dimensional photonic Lieb lattice with partial etching and elliptical holes, enabling the demonstration of a locally flat photonic band structure. This flatband spans a finite numerical aperture of $\sim$0.1 for all polarizations and $\sim$0.4 for specific polarizations. We note that the dispersionless region can be extended to even higher numerical apertures $\sim$0.6 for certain polarizations and angle of incidence by optimizing the size of the various elements of the Leib lattice as shown in Supplementary Figure 5. Furthermore, using an all-dielectric metasurface, our metasurface exhibits a high quality factor of $\sim$282 at a wavelength of around 770 nm. To further expand the dispersionless region of the two-dimensional flatband, employing circular polarization \cite{choi2025observation} to decouple the incident angle and the polarization direction can possibly present a promising avenue for future breakthroughs. This large angle/ polarization insensitive dielectric metasurface provides great opportunities as a compact filter or sensor which can be utilized in augmented reality devices (e.g., eye tracking) \cite{guenter2012foveated}.

\begin{acknowledgments}
A.M. was supported by the National Science Foundation Grant No. DMR-2019444 and NSF-210367. Part of this work was conducted at the Washington Nanofabrication Facility/Molecular Analysis Facility, National Nanotechnology Coordinated Infrastructure (NNCI) site at the University of Washington with partial support from the National Science Foundation via awards NNCI-1542101 and NNCI-2025489.
\end{acknowledgments}

\bibliography{apssamp}

\end{document}